\documentstyle[twocolumn,aps,epsf,floats]{revtex} 
\begin{document}
\draft
\preprint{}
\title{Laser induced quasicrystalline order
 in charge stabilised colloidal systems}
\author{Chinmay Das\cite{a} and H. R. Krishnamurthy\cite{b}\cite{c}}
\address{Department of Physics, Indian Institute of Science,
Bangalore 560 012, India}
\date{\today}
\maketitle
\begin{abstract}
We have studied the ordering of a two dimensional charge stabilised
colloidal system in the presence of a stationary one dimensionally
modulated laser field formed by the superposition of two modulations
with wavevectors $q_0 \tau$ and $q_0/\tau$, where $q_0$ is the
wavevector corresponding to the first peak of the direct correlation function
of the unperturbed liquid, and $\tau$ is the {\em golden mean}.
In the framework of the Landau-Alexander-McTague theory we find that
a decagonal quasicrystalline phase is stabler than the liquid or
the triangular lattice in certain regions of the phase diagram.
Our study also shows a reentrant melting phenomena for larger laser
field strengths. We find that the transition from the modulated liquid
to the quasicrystalline phase is continuous in contrast to a first order
transition from the modulated liquid to the triangular crystalline phase.


\end{abstract}               

In recent years, there has been considerable interest in  
the ordering of charge stabilised colloidal particles in the presence of
stationary laser modulations \cite{burns,ajay,acker,lasim,ladft}.
Due to their large diameter and charge ($\sim$1000e for a
particle of diameter 1000 $\AA$), 
charge stabilised colloidal particles have large polarisabilities.
The electric field of the laser beam induces dipole moments on the
colloidal particles and these dipole moments in turn interact with the
laser electric field. The resulting interaction energy of a colloidal
particle at position $r$ is equivalent to an external potential
$V_e(r) = - \frac{1}{2} \chi (E(r))^2$, where $\chi$ 
is the dielectric susceptibility
of the colloidal particles and  $E(r)$ is the electric field at $r$.
This interaction 
causes the particles to preferentially sit at the maxima and the
minima of the electric field modulations and 
thus promotes density modulations at twice the wave-vector of the
modulating electric field.
 By manipulating the field pattern
one can get complex structures, referred to as {\it Optical matter} 
\cite{burns}, for moderate field 
strengths which are easily attainable in the laboratory 
(for recent reviews on colloids and other ref.s see \cite{ajay}).
When the interaction energy with the external field is of the same
order as the thermal energy of the particles, ordering can take place
with wavevectors other than the wavevectors of the periodic 
stationary external potential $V_e(r)$.
This is due to the nonlinear coupling of the different order parameter modes.
In particular, it has been observed that when the wavevector of an
one dimensional
laser modulation is tuned to  the wavevector $q_0$ corresponding to
the first peak of  the
direct correlation function of the unperturbed liquid, the system
freezes to a triangular lattice \cite{acker}, 
and simulation studies \cite{lasim} have
also reproduced the experimental findings.
Several issues connected with this phenomenon of `` Laser Induced Freezing
(LIF)'' have been explored in a recent  Density functional theory \cite{ladft}.

Burns {\it et al.}\cite{burns} have demonstrated that a
quasi-crystalline arrangement of colloidal particles in two
dimensionally confined geometry  is obtainable  by subjecting them
to a superposition of five equiangular
coherent laser beams. In their experiment
the external field had the same symmetry of the final structure.
A two dimensional quasicrystal is characterised by density
modulations in four independent wavevectors \cite{pen}. It is
interesting to ask whether one can 
generate LIF into a quasicrystalline
structure:  i.e., have the external modulation couple directly only to
a few of the order parameter modes, and have the system
generate density modulations with the other wavevectors needed to make up
the quasicrystalline structure via the nonlinear 
coupling of the order parameter modes.

\begin{figure}[htbp]
\epsfxsize=6.5cm
\centerline{\epsfbox{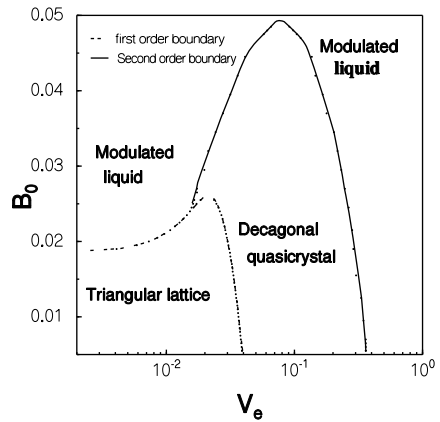}}
\caption{Phase diagram from the Landau - Alexander - McTague theory }
\label{fig:lanphase}
\end{figure}

In this work, we show using the Landau-Alexander-McTague \cite{alex} theory
that this is indeed possible, using a superposition of
two 1-d laser modulations with their
wavevectors tuned to be $q_0 \tau$ and $q_0/\tau$, where 
 $\tau (= \frac{\sqrt{5}+1}{2})$ is the  {\it golden mean}.
We show that  in the presence of such an external modulations the 
system undergoes a transition
to a {\it quasicrystalline structure} having decagonal symmetry 
for a certain range of parameters. We also 
observe a reentrant liquid phase (see figure \ref{fig:lanphase}).

The Landau-Alexander-McTague \cite{alex} theory is a mean field theory which
expresses the 
free energy(${\cal F}$) as a polynomial expansion in powers 
of the order parameters($\rho_i$)
of the system; where $\rho_i$ are the fourier components of the density,
(or better, of the molecular field) :
\begin{eqnarray}
{\cal F}& =& [\sum_{i} B_i \rho_i^2] - C [\sum_{i,j,k} \rho_i \rho_j \rho_k 
\delta_{G_i+G_j+G_k,0}]  \nonumber\\
&&+ D [\sum_{i} \rho_i^2]^2 + E[\sum_{i} \rho_i^4] 
- \sum_{i} V_e(G_i) \rho_i   \label{eqn:free};
\end{eqnarray}    

Here all the parameters $B_i, C, D\,$ and $E$ are assumed to be positive,
and $V_e(G_i)$ are fourier coefficients of the external modulation.
 To find the most stable configuration of the system 
for  a given set of parameter
values, one minimises this free energy for different choices of the order
parameter sets corresponding to different lattice arrangements. 
The quadratic term in the free energy favours the liquid phase and 
the quartic terms ensure  global
stability. When $B_i$ are sufficiently small, the cubic term clearly 
drives the system towards structures in which several 
sets of three wavevectors obey  a triangular
relationship among themselves. 
In the absence of external modulations, i.e., of  $V_e(G_i)$,
in 2-d this would favour a triangular lattice. A decagonal
quasicrystal would not be a stable phase,  because
no triangular relationships exist among the wavevectors
corresponding to a decagonal symmetry.
\begin{figure}[htbp]
\vspace{-2cm}
\epsfxsize=6.5cm
\centerline{\epsfbox{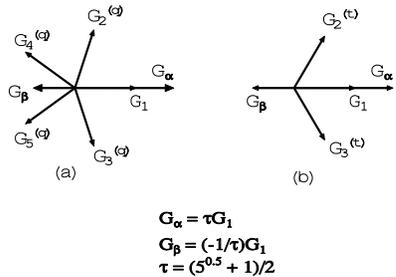}}
\caption{Density wave-vectors corresponding to {\it (a)}
modulated quasicrystal and {\it (b)} modulated triangular lattice.
Note that for each density fluctuation at wavevector {\bf G}, density 
fluctuation at {\bf - G} also is present, though not shown in the figure.}
\label{fig:lanwv}
\end{figure}

Now consider what happens when an 1-d quasiperiodic  external 
laser modulation potential of strength $V_e$  at
wave vectors $\pm G_{\alpha}$ and  $\pm G_{\beta}$ ,where
 $ G_{\alpha} = \tau G_1$ and $G_{\beta} =
-\frac{G_1}{\tau}$, and $G_1$ is any wavevector 
with magnitude $q_0$, is applied.
Our choice of wavevectors is motivated from the fact that these
wavevectors have triangular relationships with several wavevectors $G_i$
of magnitude $q_0$ and corresponding to a decagonal symmetry 
[Figure \ref{fig:lanwv}.(a)].
Noting that $\tau - \frac{1}{\tau} = 1$,
we get the following three triangular relations among the wave vectors 
defining a decagonal quasicrystal.

\begin{eqnarray}
G_1^{(q)} - G_{\alpha} - G_{\beta} &=& 0  \nonumber \\
G_2^{(q)} + G_3^{(q)} - G_{\alpha}&=& 0 \\
G_4^{(q)} + G_5^{(q)} - G_{\beta}&=& 0 \nonumber 
\end{eqnarray}

By contrast, for a triangular lattice one finds two such sets of wave-vectors 
which add to zero [figure \ref{fig:lanwv}.(b)] in
the presence of the same external field:

\begin{equation}
G_1^{(t)} - G_{\alpha} - G_{\beta} = 0; \, \, \,  
G_1^{(t)} - G_2^{(t)} - G_3^{(t)} = 0
\end{equation}

Note that all of the triangular relations for the decagonal quasicrystal
necessarily involve the wavevectors of the external modulation potential.
In contrast, only one of the two relations for the triangular lattice 
involves the modulation wavevectors.

Using the above relations, we can write down 
the free energy for a decagonal quasicrystal 
in the presence of our modulating potential as:
\begin{eqnarray} 
{\cal F}^{(q)}&=& -2 V_e [\rho_{\alpha}^{(q)}+\rho_{\beta}^{(q)}]+ \nonumber \\
&&2 B_0 [\rho_1^{(q)^2} + 2(\rho_2^{(q)^2}+\rho_4^{(q)^2})] + 
2 B_1 (\rho_{\alpha}^{(q)^2}+\rho_{\beta}^{(q)^2}) -\nonumber \\
&& 2 C [\rho_1^{(q)} \rho_{\alpha}^{(q)} \rho_{\beta}^{(q)} +
\rho_2^{(q)^2} \rho_{\alpha}^{(q)} +
\rho_4^{(q)^2}  \rho_{\beta}^{(q)}] + \nonumber \\
&&4 D [ \rho_1^{(q)^2} + 2(\rho_2^{(q)^2} + \rho_4^{(q)^2}) +
\rho_{\alpha}^{(q)^2} + \rho_{\beta}^{(q)^2}]^2 + \nonumber \\
&&2 E[\rho_1^{(q)^4} + 2(\rho_2^{(q)^4} + \rho_4^{(q)^4})
+ \rho_{\alpha}^{(q)^4} +\rho_{\beta}^{(q)^4}]
\end{eqnarray}

Here, using symmetry considerations we have set
 $\rho_2^{(q)} = \rho_3^{(q)}$ and
$\rho_4^{(q)} = \rho_5^{(q)}$.  Thus we now have five
independent order parameters for the modulated decagonal structure. 
Accordingly we get five coupled polynomial equations from
minimising the free energy with respect to 
these order parameters. 
These we have solved numerically 
using Newton's method of finding roots,
for varying values of the
parameters $B_0$ and $V_e$ and fixed values,
$B1 = 0.15$, $C = 1.0$, $D = 0.125$ and $E = 0.75$,
for the remaining parameters. 
The fixed parameter values chosen here are
the same as in \cite{acker}; except for $B_1$, which we have chosen
for convenience to be the same at $G_{\alpha}$ and $G_{\beta}$,and 
to be much larger than the maximum value considered for $B_0$.
This is motivated from the fact that the liquid structure factor 
peaks at $q_0$ and having a density modulation at some other
wave-vector will cost more energy. The qualitative features of our 
results are not sensitive to these specific values.

A similar, but separate, calculation is  done for the  triangular lattice with
two independent order parameters corresponding to triangular symmetry, 
(since, by symmetry considerations, $\rho_2^{(t)} = \rho_3^{(t)}$ )
and the remaining two order parameters corresponding to the external 
modulating field. The corresponding free energy is given by:
\begin{eqnarray}
{\cal F}^{(t)}&=& - 2 V_e [\rho^{(t)}_{\alpha} + \rho^{(t)}_{\beta} ]
+ 2 B_0 [\rho_1^{(t)^2} + 2 \rho_2^{(t)^2}] + \nonumber \\
&&2 B_1 [\rho^{(t)^2}_{\alpha} + \rho^{(t)^2}_{\beta} ] -
2 C [\rho^{(t)}_1 \rho^{(t)^2}_2 + 
\rho^{(t)}_1 \rho^{(t)}_{\alpha} \rho^{(t)}_{\beta} ] + \nonumber \\
&&4 D[\rho^{(t)^2}_{\alpha} + \rho^{(t)^2}_{\beta} + 
\rho_1^{(t)^2} + 2 \rho_2^{(t)^2}]^2  + \nonumber \\
&&2 E[\rho^{(t)^4}_{\alpha} + \rho^{(t)^4}_{\beta} + \rho_1^{(t)^4} + 
2 \rho_2^{(t)^4} ]
\end{eqnarray}

\begin{figure}[htbp]
\epsfxsize=6.5cm
\centerline{\epsfbox{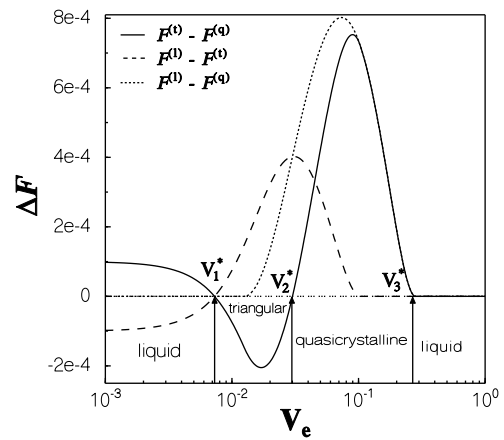}}
\caption{Free energy difference for $B_0=0.02$.
The free energy which is lower than others corresponds to
the stable phase at a particular external field.}
\label{lanop1}
\end{figure}

In Newton's method for root finding, when multiple solutions, corresponding
to multiple minima (more generally extrema) of the free energy, are 
present, the solution to which the result converges 
depends on the initial inputs for the order parameters. 
In our calculations each of the two free energy functions
corresponding to the triangular lattice and the quasicrystalline structure
was effectively minimised with the
initial guess value for the induced order parameters being 0.0 or 0.5.
We found that the results converged to one of the two solutions:
one corresponding to the modulated liquid phase, where only
$\rho_{\alpha} (= \rho_{\beta})$ and $\rho_1$ 
are nonzero, and the other to the modulated crystalline
or the quasicrystalline phase, where the other order parameters 
are also nonzero. Typically, a maximum 20000 iterations were
carried out in Newton's method for finding roots.
It was checked that the results were not dependent on the initial guess
values except for the flow to one of the two solutions alluded to above.
For each parameter set the phase corresponding to the lowest 
free energy amongst the different solutions was chosen as the stable phase.

\begin{figure}[htbp]
\epsfxsize=6.5cm
\centerline{\epsfbox{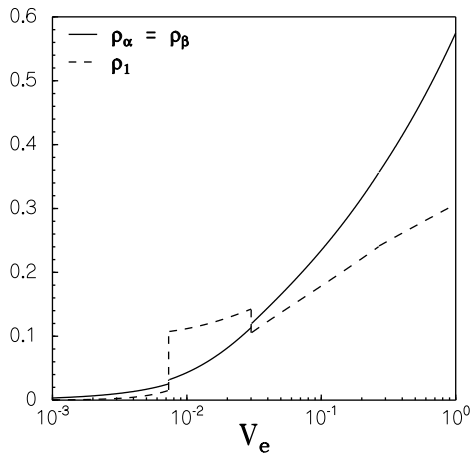}}
\caption{Results for the order parameters 
($\rho_{\alpha} = \rho_{\beta})$, 
directly coupling 
to the external field 
and  for the order parameter $\rho_1$ with wavevector 
$G_1$ parallel to the wavevector of the external modulation for $B_0=0.02$.
}
\label{lanop2}
\end{figure}

\begin{figure}[htbp]
\epsfxsize=6.5cm
\centerline{\epsfbox{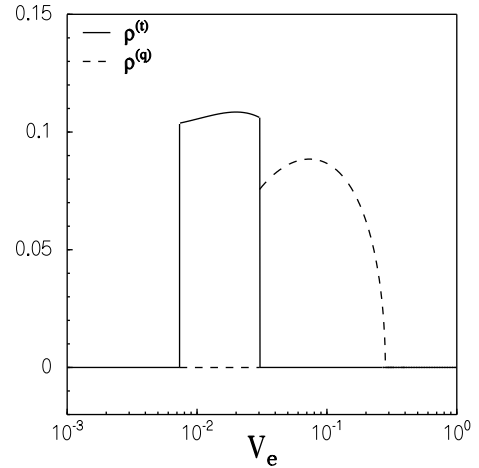}}
\caption{Order parameters corresponding to triangular order 
($\rho^{(t)} = \rho_2^{(t)} = \rho_3^{(t)}$)
and those corresponding to quasicrystalline order 
($\rho^{(q)} = \rho_2^{(q)}  = \rho_3^{(q)} =\rho_4^{(q)} = \rho_5^{(q)}$) for
$B_0=0.02$. }
\label{lanop3}
\end{figure}

In Figure \ref{lanop1} we present results for the 
free energy differences for the three different kinds of order 
as a function of $V_e$  for $B_0=0.02$. From these 
the phase with the lowest free energy is easily chosen, and the corresponding 
order parameters are shown in figures \ref{lanop2} 
and \ref{lanop3}
as functions of $V_e$.

The full phase diagram is shown in figure \ref{fig:lanphase} . We term
the phases as modulated, since the order parameters with wavevectors 
along the wavevectors of the external modulation
are typically higher than those with wavevectors in the other directions.

Depending upon the value of $B_0$,
as the external field is increased from zero, several interesting
transitions are discernible from the phase diagram. 
For low values of $B_0$ and of the modulation potential, 
the system is a triangular lattice. As the field strength is
increased, the system undergoes a first order phase transition to
the modulated decagonal quasicrystal.
At still higher field values it goes to the 
modulated liquid phase via a continuous transition \cite{artifact}.

For $B_0$ between 0.02 and 0.025, with increasing field strength
one encounters a first order LIF transition from the modulated liquid to
the triangular crystal; a first order transition from the triangular crystal
to the decagonal quasicrystal and finally a continuous transition to the 
modulated liquid reentrant phase from the quasicrystalline phase 
(fig. \ref{lanop3}).

It is worth pointing out that near $B_0 = 0.025$ 
one can go from the modulated liquid directly
to the quasicrystalline phase via a continuous LIF transition and thence
to the triangular phase via a first order transition. With increasing field
the system goes once again to the quasicrystalline phase through a
reentrant first order transition. 
At still higher fields the transition to a reentrant liquid phase
occures via a second order phase boundary. 

For $B_0 > 0.0265$
the system goes into the quasicrystalline order from the liquid 
phase via a continuous transition as the
field strength is increased and then again melts to the modulated liquid 
phase at still higher field.
Above $B_0=0.05$ the system remains in the modulated
liquid phase for all field strengths.

The fact that at low external fields, the system either freezes to a
triangular lattice or remains a liquid depending upon the value of
$B_0$  and that the transition is first order
is consistent with our knowledge about colloidal systems
in the absence of any external field modulation.

At moderate field strengths, the system gains  energetically by
having density modulations corresponding to the wave-vectors of
the modulating laser field. At higher field strengths, even though the 
external field modulation is 1-dimensional, density modulations
corresponding to a 2-dimensional triangular or decagonal symmetry develop 
because of the nonlinear coupling among the order parameter modes.

At still higher fields the continuous melting is consistent with
previous simulations \cite{lasim} and the Landau-Alexander-McTague meanfield 
analysis of the laser induced
freezing \cite{acker,artifact}. 

The boundary between the triangular lattice and the quasicrystalline
phase is first order because the two structures are of completely 
different symmetry and one can not  deform a triangular lattice 
continuously to get a decagonal symmetry. 

In contrast, the phase boundary
between the modulated liquid and the crystalline and quasicrystalline
phases can be first order or continuous. The mechanism that determines
which has been discussed in detail in ref \cite{ladft}. Basically,
in the modulated liquid phase $\rho_{\alpha} = \rho_{\beta}\,$ and $\,\rho_1$ are
nonzero. Now consider setting up a Landau expansion for the free energy 
in powers only of the additional order parameters that characterise 
the crystalline or quasicrystalline phases ( i.e., with 
respect to the modulated liquid phase). Such an expansion has only
even order invariants in the additional order parameters  for the
cases we are discussing. The order of the transition depends on the 
signs of $T_2$ and $T_4$, the second order and fourth order coefficients  
respectively, in such an expansion. A first order transition ensues
when $T_4 < 0$; whereas a continuous transition results when $T_4 > 0$,
the phase boundary being determined by the condition $T_2 = 0$.

When the field strength is very large, the linear term coupling the
laser modulation potential with the order parameters
at the modulation  wavevectors  
seems to be the most dominant term
for lowering the free energy.And we find that
having density modulations along other
wavevectors no longer lowers the free energy. So the decagonal phase melts
to give the reentrant modulated liquid \cite{artifact}.

In conclusion we have shown that ,by subjecting a 2-dimensionally
confined charge-stabilised colloidal liquid to  a superposition of
two 1-d laser modulations with their
wavevectors tuned to be $q_0 \tau$ and $q_0/\tau$ (where $q_0$ is
the wavevector of first peak of the liquid structure factor),  
one can generate laser-induced-freezing into a decagonal
quasicrystalline order. We have also shown  that for larger
laser field strengths, within mean field theory this transition 
is continuous and  shows a reentrant melting back to the modulated
liquid phase. It would be of interest if the experiments of ref.
\cite{acker} can be extended to explore these transitions
and the resulting quasicrystalline phase.

We thank A.K.Sood, T. V. Ramakrishnan, S. Ramaswamy, 
R. Pandit and Rangan
Lahiri for many useful discussions. One of us (CD) thanks
CSIR, India for support.


\begin{thebibliography}{15cm}
\bibitem{fifth}
One can stabilise decagonal quasicrystalline order in two dimensions for a 
single component system  \cite{bak}, even in the absence of any external
modulations, by  adding a large  fifth order term in 
equation(\ref{eqn:free}) . But there is no physical reason 
why the fifth order term will be
more significant than the cubic term. Since we are specifically
interested in quasicrystalline order induced by external modulations,
in our analysis we keep only terms
upto fourth order in the order parameters. Inclusion of a small 
fifth order term will
only make the quasicrystalline phase more stable, without changing
the qualitative features of our phase diagram.
\bibitem{artifact}
As discussed in  ref.\cite{ladft} this last, melting 
transition, and the stability
of the modulated liquid phase for low $B_0$ and arbitrarily large $V_e$,
are probably  artifacts of the 
Landau- Alexander-Mctague theory.
\end{thebibliography}
\end{document}